\documentclass[twocolumn,showpacs,preprintnumbers]{revtex4}
\usepackage{graphicx}
\usepackage{dcolumn}
\usepackage{bm}
\usepackage{color}

\graphicspath{%
    {converted_graphics/}
    {/}
}
\begin{document}

\title{Toward a test of angular momentum coherence in a twin-atom interferometer}

\author{Carlos R. de Carvalho$^{1 (a)}$ , Ginette Jalbert$^{1 (b)}$, Fran\c{c}ois Impens$^{1 (c)}$, J. Robert$^{2}$, \\ Aline Medina$^{3}$, F. Zappa$^{4}$, and N. V. de Castro Faria$^{1}$} 
\affiliation{
$^1$Instituto de F\'\i sica, UFRJ, Cx.~Postal 68528, Rio de Janeiro, RJ 21941-972, Brazil
\\
$^2$Laboratoire Aim\'{e} Cotton CNRS, Univ Paris Sud-11, 91405 Orsay Cedex, France
\\
$^3$Departamento de F\'\i sica, UFBA, Salvador, BA 40210-340, Brazil
\\
$^4$Departamento de F\'\i sica, UFJF, Juiz de Fora, MG 36036-330, Brazil
}



\begin{abstract}
We present a scheme well-suited to investigate quantitatively the angular momentum coherence of molecular fragments. Assuming that the dissociated molecule has a null total angular momentum, we investigate the propagation of the corresponding atomic fragments in the apparatus. We show that the envisioned interferometer enables one to distinguish unambiguously a spin-coherent from a spin-incoherent dissociation, as well as to estimate the purity of the angular momentum density matrix associated with the fragments. This setup, which may be seen as an atomic analogue of a twin-photon interferometer, can be used to investigate the suitability of molecule dissociation processes -- such as the metastable hydrogen atoms H($2^2 S$)-H($2^2 S$)  dissociation - for coherent twin-atom optics. 
\end{abstract}


\maketitle

In spite of the innumerous experiments confirming the predictions of Quantum Mechanics with an excellent accuracy, the physical foundations of this theory have been questioned since its early days. The milestone papers by Einstein, Poldosky and Rosen \cite{Einstein35}  and  Schr\"odinger \cite{Schrodinger35}, grounded a so-far epistemological discussion~\cite{Bohr58} into the experimental reality thanks to the formulation of the EPR paradox. Later, Bohm provided a {\it gedankenexperiment} suitable to test the EPR paradox~\cite{Bohm89}. As pictured by Bohm, such experiment would use a pair of atoms, corresponding to a pair of spin-$(1/2)$ particles coming from a molecular fragmentation, in order to probe the existence of non-classical correlations between the particles spins. 
In the late 60s Clauser {\it et al.} \cite{Clauser69} transposed Bell's analysis of the {\it gedankenexperiment} of Bohm~\cite{Bell} to systems involving photons instead of massive particles. Finally, nearly fifty years after the initial formulation of the EPR paradox, the quantum non-locality was demonstrated in the optical domain thanks to the celebrated experiments of Aspect et al.~\cite{Aspect}. 

It is nevertheless appealing to go back to Bohm's original idea of testing non-classical correlations with the spin observables of massive particles. Bohm's main concern  was the maintenance of the spin coherence between the spin-$(1/2)$ massive particles, an issue later discussed by Englert, Schwinger and Scully~\cite{Schwinger88}, and more recently by Oliveira and Caldeira~\cite{Caldeira06}. In order to conduct such fundamental tests of quantum mechanics~\cite{Fry00}, it is essential to have a spin-coherent twin-particle source.\\

Beyond the tests of quantum non-locality, the development of twin-atom interferometry offers particularly exciting perspectives in atom optics~\cite{CroninRMP09}.  This field has reached a state of the art enabling the observation of basic non-linear phenomena such as the four-wave mixing or quantum phenomena such as the Hanbury-Brown Twiss effect with matter waves~\cite{Deng99,WestbrookNature07}. Besides, there has been a considerable effort and enthusiasm around EPR pairs with atoms, see for instance the work done by Hagley {\it et al}~\cite{Hagley97}, in the context of cavity QED, and  the work by Tanabe {\it et al}~\cite{Tanabe09} in atom and molecule physics, and Bucker {\it et al} in Bose-Einsten Condesate~\cite{Bucker11}. Undeniably, the realization of reliable twin-photon sources with correlated angular momenta~\cite{Zeilinger} has become a key component of experimental quantum optics. Similarly, the obtention of twin-atom sources preserving the angular momentum coherence would be a significant step forward in quantum atom optics. One may then observe, with massive atoms, quantum effects specific to the pair production of indistinguishable particles, and exploits the entanglement in the degrees of freedom (dofs) resulting from energy and angular momentum conservation. The coherence between the angular momentum states of the particle is, again, a prerequisite in order to observe entanglement in such atom interferometers.\\

In this paper, we present a  scheme based on a symmetric double Stern-Gerlach atom interferometer, which is suitable to estimate quantitatively the angular momentum coherence of the fragments issued from a molecular dissociation. In spite of the several Stern-Gerlach atom interferometers accomplished to date~\cite{Robert-Baudon,SGA07}, to our knowledge, no configuration able to treat this basic issue, concerning the spin coherence between two atoms, has been done neither proposed. 

Previous experimental work accomplished by some of us on the excitation and dissociation of molecular hydrogen~\cite{Medina11,Medina12} and on the existence of the H($2^2 S$)-H($2^2 S$) dissociation channel~\cite{Robert13} strongly advocate for the pair H($2^2 S$)-H($2^2 S$) of excited metastable states as a candidate for a twin-atom source. An important point is that the state of the art enables an accurate control of the internal degrees of freedom associated with the hyperfine structure of these states~\cite{Robert-Baudon,Lamb}. The setup considered here uses extensively a firmly established experimental technique, namely the Stern-Gerlach atom interferometry, particularly well-suited to manipulate H($2^2 S$) atomic beams~\cite{Robert-Baudon}. The present scheme, grounded by a strong experimental knowledge of the manipulation of metastable states H($2^2 S$), may actually be extended to other atomic systems. Such extension is, however, conditioned to the existence of long-lived atomic fragments and to an efficient control of their internal degrees of freedom -� desirable features which are indeed guaranteed for the metastable H($2^2 S$) states.

\begin{figure}[htbp] 
\begin{center}
  \includegraphics[width=8.8cm]{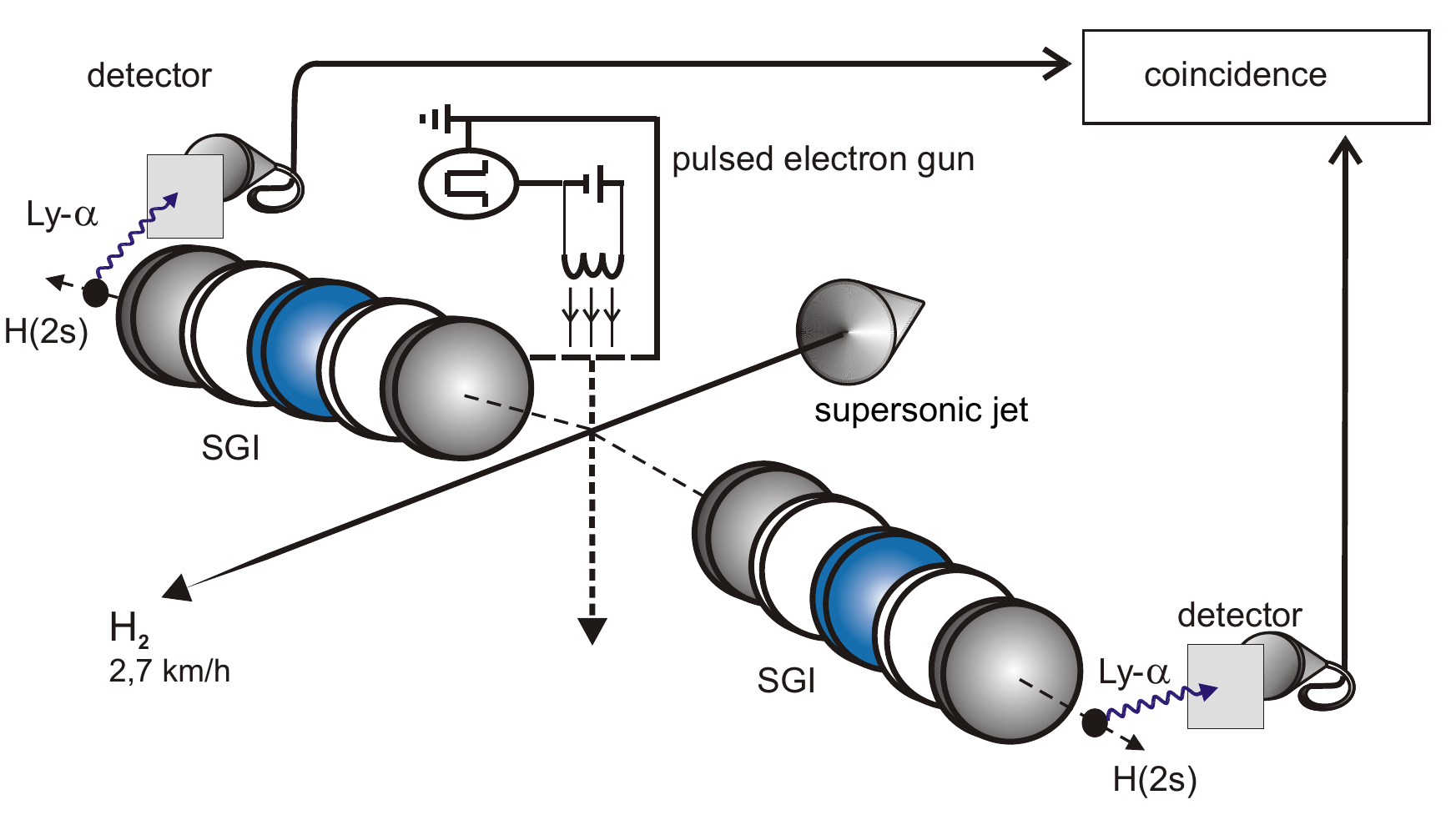}
  \caption{  A H$_2$ beam produced in a supersonic jet is bombarded by electron pulses discharged from a pulsed electron gun. The desired atomic fragments are the pair H($2^2 S$)-H($2^2 S$) coming from the dissociation of the excited H$_2$ molecules, which remain in the state $|F_T=0, M_{F_T} =0\rangle$. Each arm contains a Stern-Gerlach interferometer (SGI).} 
  \end{center}
  \label{FigA}
\end{figure}

In the discussion to follow, the scheme proposed combines the elements of our {\it time-of-flight} spectroscopy setup~\cite{Medina11,Medina12} with the coincidence experiment~\cite{Robert13} and the single Stern-Gerlach atom interferometer. Even though the double Stern-Gerlach atom interferometer configuration is new to our knowledge, each branch of our double interferometer - on Fig.~\ref{FigA}- has exactly the design already used successfully   by Robert and co-workers \cite{Robert-Baudon}.  In addition, one assumes that two atomic fragments, with a quantum number for the norm   of the angular momentum equal to one, come out from the dissociation of an excited molecule with a null total angular momentum. 

In usual experimental conditions, the atomic fragments leave the collision region with almost opposite velocities much larger than the initial molecule CM velocity.   As a direct consequence of our assumption of null initial total angular momentum, atomic fragments leave the collision zone carrying away an opposite magnetic moment. Thus, one should design the double Stern-Gerlach interferometer as to select opposite projections of the transverse/longitudinal angular momentum on both sides.

In the discussion below, we shall use a single set of axis $O_x,O_y,O_z$ to quantize the angular momentum in the whole experiment, independently of the local magnetic field orientation. Should we select the same angular momentum projections on the right and on the left, only atoms with a null angular momentum projection could possibly reach simultaneously the detectors. This would prevent us from observing any possible interference in the coincidence detection of the fragments. This is why the left side of the experiment uses a magnetic field of reversed direction as compared to the right side - the proposed system is indeed invariant by the symmetry $\mathbf{r} \rightarrow - \mathbf{r}$ with respect to the collision center.

\begin{figure}[htbp] 
\includegraphics[width=8.5cm]{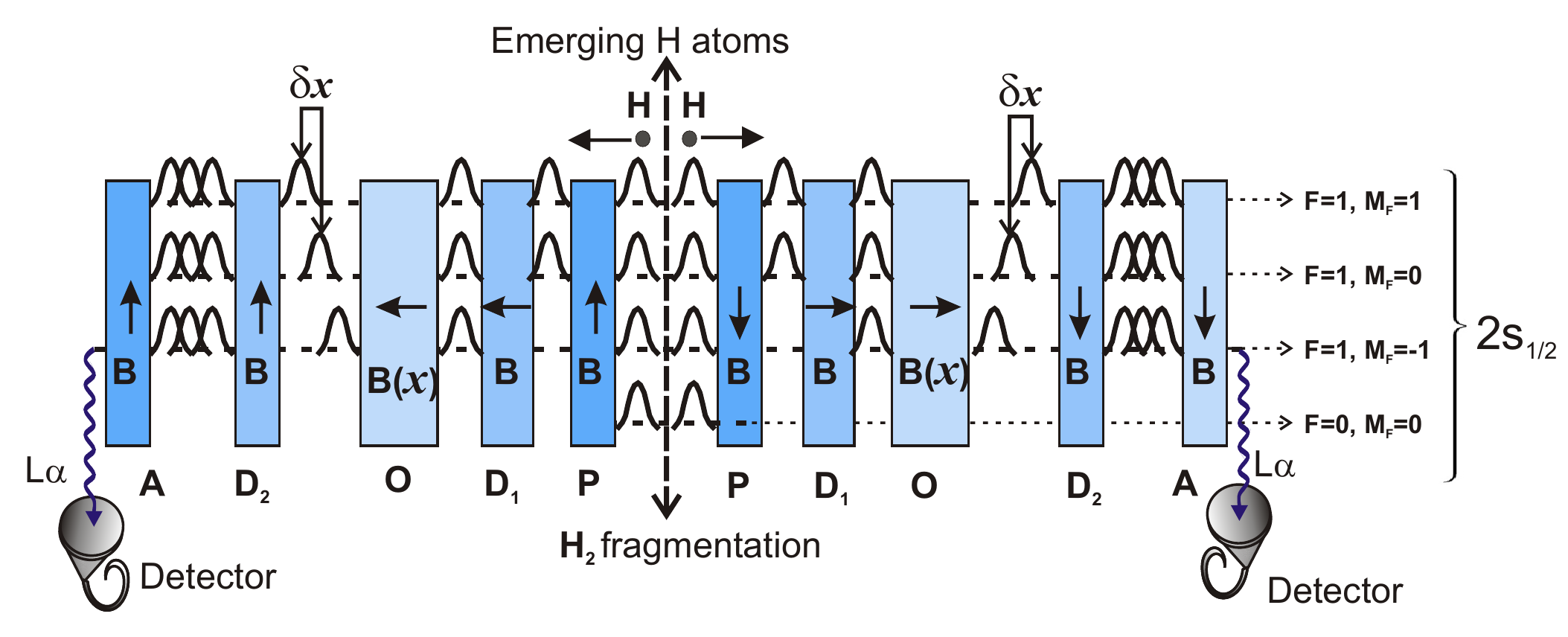}
\caption{(color  online). Double Stern-Gerlach atom interferometer. The measurement apparatus (in the center of mass frame): the atomic polarizers {\bf P}, which select the hyperfine structure states
   $\{ | f=1, m_z= 1 \rangle ,| f=1, m_z=0 \rangle \}$ on the right [  $\{ | f=1, m_z= -1 \rangle ,| f=1, m_z=0 \rangle \}$  on the left]; the beam-splitters {\bf D}  which split each of these states into a linear combination of the $| f=1, m_x=\pm1,0 \rangle$ states, the phase objects {\bf O}, which induce a phase shift in each of the hyperfine structure states; the analyzers {\bf A}, which provoke a resonant decay of one of the $m_z= -1  $ states on the right [$m_z=+1$ on the left] providing the detection signal. The arrows inside the boxes indicate the magnetic field direction in each device.} 
\label{fig:doubleSGAinterferometer}
\end{figure}

The double atom interferometer depicted on Fig.~\ref{fig:doubleSGAinterferometer} consists in two equivalent arms, each one containig a Stern-Gerlach interferometer (SGI). 
Let us first consider the right side of the experiment. It is known that the atoms H($2^2 S$) have an hyperfine structure of total angular momenta $| f=0; m_z =0 \rangle$, and $| f=1; m_z =1, 0, -1 \rangle$.
A properly tuned magnetic field permit indeed the elimination of the atomic states $| f=0; m_z =0 \rangle$ and $| f=1; m_z = -1 \rangle$  upon atomic passage in a polarizer~\cite{Lamb}. Initially, one sends the atomic beam into such a transverse polarizer eliminating atoms with a transverse angular momentum $m_z =-1 $, and thus implementing a projection operator on the angular momentum subspace $\{ | f=1, m_z =1 \rangle, | f=1, m_z =0 \rangle  \} $. 

This device is followed by a magnetic field gradient operating as a phase object, which induces a time delay between atomic wave-packets attached to different longitudinal magnetic moments $m_x = +1, 0, -1$.  It is essential that the direction magnetic field switches abruptly from vertical ($O_z$) to horizontal ($O_x$)  between the polarizer and the phase object, so that  the atomic quantum state is not altered between these devices. Such abrupt change can be seen as the atomic analogue of an optical beam-splitter, since it splits each atomic beam of definite $ m_z $ into a superposition of three modes $m_x= \pm 1, 0$ by projecting the quantum state on a new basis. These modes then acquire longitudinal separation under propagation of the longitudinal magnetic field of the phase object. A second  abrupt change immediately after the phase object, putting
back the magnetic field in the vertical $O_z$ direction, plays the role of a recombining beam-splitter. Last, an analyser provides us with a signal - obtained thanks to a Zeeman quenching~\cite{Robert89}- proportional to the atomic population with a transverse magnetic moment $m_z=-1$. On the left side, one first eliminates the atoms with an angular momentum $m_z=+1$. The orientation of the magnetic field along the other arm is reversed, and finally after recombination in the last beam-splitter, the analyser provides a signal proportional to the atomic population with a quantum number $m_z=+1$.

We now proceed to the analysis of the atomic propagation in the envisioned system, working in the  Schr\"odinger picture.  The system may be described through a global unitary evolution operator of the form:
\begin{equation}
\hat{U} = \mathcal{N} \left[  \hat{O}  \hat{U}_{E}(t_2,t_1)     \hat{P} \hat{U}_{E}(t_1,t_0) \frac {} {} \right] \label{eq:general evolution Schrodinger picture}
\end{equation}
where the operators $ \hat{P}$ and $\hat{O}$ capture the action of the polariser and the phase object respectively, present in each arm of the interferometer. The operators $\hat{U}_{E}(t,t')$ account for the external atomic motion, $t_0$ is the initial time immediately after the dissociation and the instants $t_1,t_2$ correspond to the entrance time of the atoms in the successive devices.  $\mathcal{N}$ refers to the normalization of the quantum state, necessary here because of the projective measurements operated by the polarizer and analyser. While the duration of the atomic motion in these elements may be taken as very short, this assumption is no longer legitimate for the propagation in the phase object, which operates a spatial splitting of the atomic wave-packets thanks to a necessarily finite propagation time. However,  this propagation may be treated as an effective instantaneous interaction in the spirit of the $ttt$ scheme for atom interferometry~\cite{tttscheme}. The action of the external atomic evolution $\hat{U}_{E}(t,t') $ on the atomic wave-packets is well-known and given by the atom-optical $ABCD$ propagation method~\cite{ABCD}. For sake of simplicity, in the discussion to follow, we shall deliberately ignore the effects of the external atomic motion in-between the different stages, assuming that the successive interferometer elements are very close one to another. This assumption avoids several technicalities, and permits us to focus the discussion on the basic physical principles at work in the experiment. 
 
We now explicit the operators involved in Eq.(\ref{eq:general evolution Schrodinger picture}). As discussed above, we simply treat the operators $\hat{U}_{E}(t,t') $ as identity operators. Since the polarizers on each side filter different angular momentum states, the corresponding operators actuate simultaneously on the external and internal atomic dofs, namely $ \hat{X}=\hat{X}_1 \otimes \hat{X}_2$  for $ \hat{X}=\hat{P},\hat{O}$ with 
 \begin{eqnarray} 
 \hat{P}_j  & = &    | 0_j \rangle \langle  0_j | +   \Theta(x_j)  | 1_j \rangle \langle 1_j  |+  \Theta(-x_j)   | -1_j \rangle \langle -1_j  | \nonumber \\
\hat{O}_j & = &   \Theta(x_j)  e^{\frac i \hbar \hat{p}_{x \: j}  \hat{F}_{x  j} \hat{\Delta} x_R}
+ \Theta(-x_j)  e^{\frac i \hbar \hat{p}_{x \: j}  \hat{F}_{x  j} \hat{\Delta} x_L}  \label{eq:operators P O}
\end{eqnarray}
 where $j=1,2$ labels the atom.  $\Theta(x)$ is the Heaviside step function such that $\Theta(x)=1$ if $x > 0$ and $\Theta(x)=0$ if $x \leq 0$. We have used the short-hand notation $| m_1 m_2 \rangle \equiv | f_1=1; m_{z 1} = m_1 \rangle \otimes | f_2=1; m_{z 2} = m_2 \rangle $ to denote the angular momentum states, and introduced the longitudinal momentum operator  $\hat{p}_{x \: j} $ for the $j_{\rm th}$ atom. The operator $\hat{F}_{x  j}$, acting on the $j_{\rm th}$ particle, corresponds to the angular momentum projection along the $O_x$ axis. We discuss below how its connection to the angular momentum operator $\hat{F}_{z j}$ along $O_z$ induces an atomic beam-splitting.  \\
 
 The Stern-Gerlach imprints a phase shift on the atomic waves, which depends on the longitudinal magnetic moment, and on which side of the experiment on the atomic wave propagate. More precisely, the shifts in atomic position imprinted on the right and left hand side depend on the longitudinal kinetic energy $\hat{E}_{x \: j}= \frac {\hat{p}_{x \: j}^2} {2 m}$ and on the local magnetic field through  $\hat{\Delta} x_R = (m_A  g \mu_B) / (2 \hat{E}_{x \: j}) \int_{x_1}^{x_2}  B( x) dx  $ and $\hat{\Delta} x_L = (m_A  g \mu_B)/ ( 2 \hat{E}_{x \: j}) \int_{-x_2}^{-x_1}  B( x) dx $ with $m_A$ the atomic mass, and where the phase object zone is defined for  $x_1 < x < x_2$ on the right and $-x_2 < x < -x_1$ on the left. Faster atomic waves are less influenced by the longitudinal magnetic field since they propagate during a shorter time in the phase object. In practice, we neglect dispersion effects in the Zeeman phase acquired by the atomic waves, i.e. one takes $\hat{p}_{x \: j} \equiv p_0$ for right-propagating atomic waves, and $\hat{p}_{x \: j} \equiv - p_0$ for left-propagating atomic waves in the previous expressions.  Here, we plan to use more specifically an antisymmetric magnetic field profile $B(-x)=-B(x)$. It is essential to note that the eigenstates of the polarizers - of definite transverse angular momentum $m_{z j}$ - are not eigenstates of the longitudinal phase object, for which the relevant quantum number is the longitudinal magnetic moment- $m_{x j}.$ This switch of polarization axis, at the origin of the desired beam-splitting, is expressed by the relation $\hat{F}_{x \:j}= \hat{D}_j(-\pi/2) \hat{F}_{z \: j}  \hat{D}_j(\pi/2)$ where $\hat{D}_j(\pi/2)$ operates a rotation of angle $+\pi/2 $ of the angular momentum quantization axis (from $O_z$ to $O_x$) of the $j_{\rm th}$ particle and  $\hat{D}_j(-\pi/2)$ operates the inverse transform~\cite{Varshalovich}. Note that, by virtue of the relation 
 \begin{equation}
 e^{i \frac {p_0} {\hbar}  \hat{F}_{x  j} \Delta x_{R,L}}= \hat{D}_j(-\pi/2)  e^{i \frac {p_0} {\hbar}   \hat{F}_{z  j}  \Delta x_{R,L}} \hat{D}_j(\pi/2) \,,
 \end{equation}
  the operator $\hat{U}$ defined in Eq.~(\ref{eq:general evolution Schrodinger picture}) can be expressed as a product of operators involving the angular momentum in a single direction ($O_z$ for instance) and of operators $\hat{D}_j(\pm \pi/2)$ relating the $O_x$ and $O_z$ angular momentum basis.

We now discuss the form of the initial atomic density matrix, which reflects the spin coherence of the molecule dissociation. Since the atomic fragments are produced by pairs of indistinguishable particles - bosons in the case of $H$ atoms -, in the analysis of the atomic propagation, one should consider only 
 probability amplitudes between properly symmetrized quantum states. One assumes that the external atomic wave-function is symmetric, of the form $ \langle x_2, x_1 | \psi_E^{\rm out}  \rangle  = F(x_1,x_2)+ F(x_2,x_1)$ with $F(x,x') =  (A / 2 |x-x'|)  \Theta(x-x' )e^{ik (x-x')  } $ and $A$ a normalization constant chosen as a real number without loss of generality. This external state is based on the continuum vibrational state of the excited molecule for the two nuclei~\cite{Siebbeles91}  $\psi_E(\mathbf{r}_1,\mathbf{r}_2) = (A' / | \mathbf{r}_1-\mathbf{r}_2 |) \sin (k |\mathbf{r}_1-\mathbf{r}_2| + \delta)$, where only the outgoing spherical waves have been retained. Note that the phase $\delta$ is reminiscent of the excited molecular state and contains relevant information about the shape of the repulsion potential between the nuclei. Here, our assumption that the molecule has no rotational energy makes the interferometer insensitive to this phase.  Immediately after the molecule dissociation, given  the null total angular momentum,  the twin particle indistinguishability, and the symmetry of the external wave-function, the space of acceptable angular momentum states is two-dimensional, spanned by the states $| \psi_0 \rangle =| 0_1 0_2 \rangle $ and $| \psi_1 \rangle  =  \frac {1} {\sqrt{2}} \left( | 1_1 -1_2 \rangle +  | -1_1 1_2 \rangle \right)$. The states $| \psi_0 \rangle$ and $| \psi_1 \rangle$ correspond to atomic fragments carrying respectively an angular momentum of either null or opposite projection along the axis $O_z$. Should the particles be distinguishable, the sub-space of acceptable angular momentum states would be spanned by the three dimensional basis $\{ | 0_1 0_2 \rangle, | 1_1 -1_2 \rangle, | -1_1 1_2 \rangle\} $ instead.

The atomic fragments thus behave as an effective two-level system, whose coherence may be characterized by a Bloch vector. For a spin-coherent molecule dissociation, the quantum state of the system immediately after the dissociation would be  $| \Psi_S \rangle = |   F =0, M_{z} =0 \rangle $ (where $F$ and $M_{z}$ are the quantum numbers for the total angular momentum), or according to a Clebsh-Gordan decomposition~\cite{Varshalovich} as $
| \Psi_S \rangle= -  \sqrt{\frac {1} {3}} | \psi_0 \rangle + \sqrt{\frac {2} {3}} | \psi_1 \rangle$ up to a global phase.  This corresponds to a Bloch vector pointing on the Poincar\'e sphere. More generally, for a partially coherent dissociation, the angular momentum dofs are described by an initial density operator of the form  

\begin{equation}
\hat{\rho}^0_S = \frac 1 3 | \psi_0 \rangle \langle \psi_0| + \frac {2} {3}  | \psi_1 \rangle \langle \psi_1 | - \lambda \frac {\sqrt{2}} {3}  \left(  | \psi_0 \rangle \langle \psi_1 | +  | \psi_1 \rangle \langle \psi_0 | \right) \label{eq:initial density matrix}
\end{equation}
with $0 \leq \lambda \leq 1 $. The parameter $\lambda$ quantifies the spin coherence, indeed $\lambda=1$ and $\lambda=0$ correspond respectively to a completely spin-coherent and completely spin-incoherent dissociation. In what follows, we show the dependence of the interference pattern -- obtained by  coincidence detection after the double SG interferometer -- 
with respect to the parameter $\lambda$. Indeed, this parameter is directly connected to the purity of the angular momentum density matrix through $\gamma={\rm Tr} \hat{\rho}^2=5/9+4/9 \lambda^2$ and to the corresponding linear entropy~\cite{Kwiat04} $S=1-{\rm Tr} \hat{\rho}^2=4(1-\lambda^2)/9.$\\

With the Eqs.(\ref{eq:general evolution Schrodinger picture},\ref{eq:operators P O}) at hand, it is  a straightforward task to derive the operation of our system onto a coherent (or incoherent) superposition of atomic states. The final density matrix $\hat{\rho}_F$ is indeed given by $\hat{\rho}_F = \hat{U} \left[ \frac {} {}  \hat{\rho}^0_S \otimes  | \psi_E^{\rm out}  \rangle  \langle  \psi_E^{\rm out} | \frac {} {}  \right]   \hat{U}^{\dagger}   $.  The detection apparatus provides a signal proportional to the counting rates of the coincidence measurements of the atoms possessing an angular momentum component $m_z=+1$ on the right and $m_z=-1$ on the left. Formally, the detection signal can be analysed in terms of the projection of the density matrix onto the symmetrized quantum state 
\begin{eqnarray}
| \psi^{\rm det} \rangle  & = & \frac {1} {\sqrt{2}} \left( | x_1= -D/2 \rangle  | 1_1 \rangle  | x_2=D/2 \rangle | -1_2 \rangle \right. \nonumber \\
&+ & \left. | x_1= D/2 \rangle |  -1_1 \rangle | x_2= D \rangle |  1_2 \rangle  \right) \,, \nonumber 
\end{eqnarray} namely the counting rate is proportional to $I_{12} = \langle \psi^{\rm det} |   \hat{\rho}_F    | \psi^{\rm det} \rangle $, and is given by
\begin{eqnarray}
I_{12} & = & \mathcal{C} \sin ^2  \frac {\phi_L} {2} \sin ^2 \frac {\phi_R}{2} \left[ \frac {} {} 4 \lambda  \sin \phi_L \sin \phi_R \right. \nonumber \\
& & \left. +\cos \phi_L (5 \cos \phi_R +3)+3 \cos \phi_R+5  \frac {} {} \right] \nonumber \\ \label{eq:detection signal}
\end{eqnarray}
where $\phi_L=  \frac {p_0}  {\hbar}  \Delta x_L $  ($\phi_R=  \frac {p_0}  {\hbar}  \Delta x_R $)  are the atomic phases imprinted by the left (right)-side magnetic field on the atoms. $\mathcal{C} $ is a constant accounting for the value of the twin-particle external wave-function, as well as for the partial efficiency of the detection system and for the normalisation factor $\mathcal{N}$ in Eq.(\ref{eq:general evolution Schrodinger picture}). To analyze the system coherence, one sets the magnetic field on the left side - thus determining $\phi_L$ - and scans the phase $\phi_R$ on the interval $[0,2 \pi]$.

\begin{figure}[tbp] 
\includegraphics[width=7 cm]{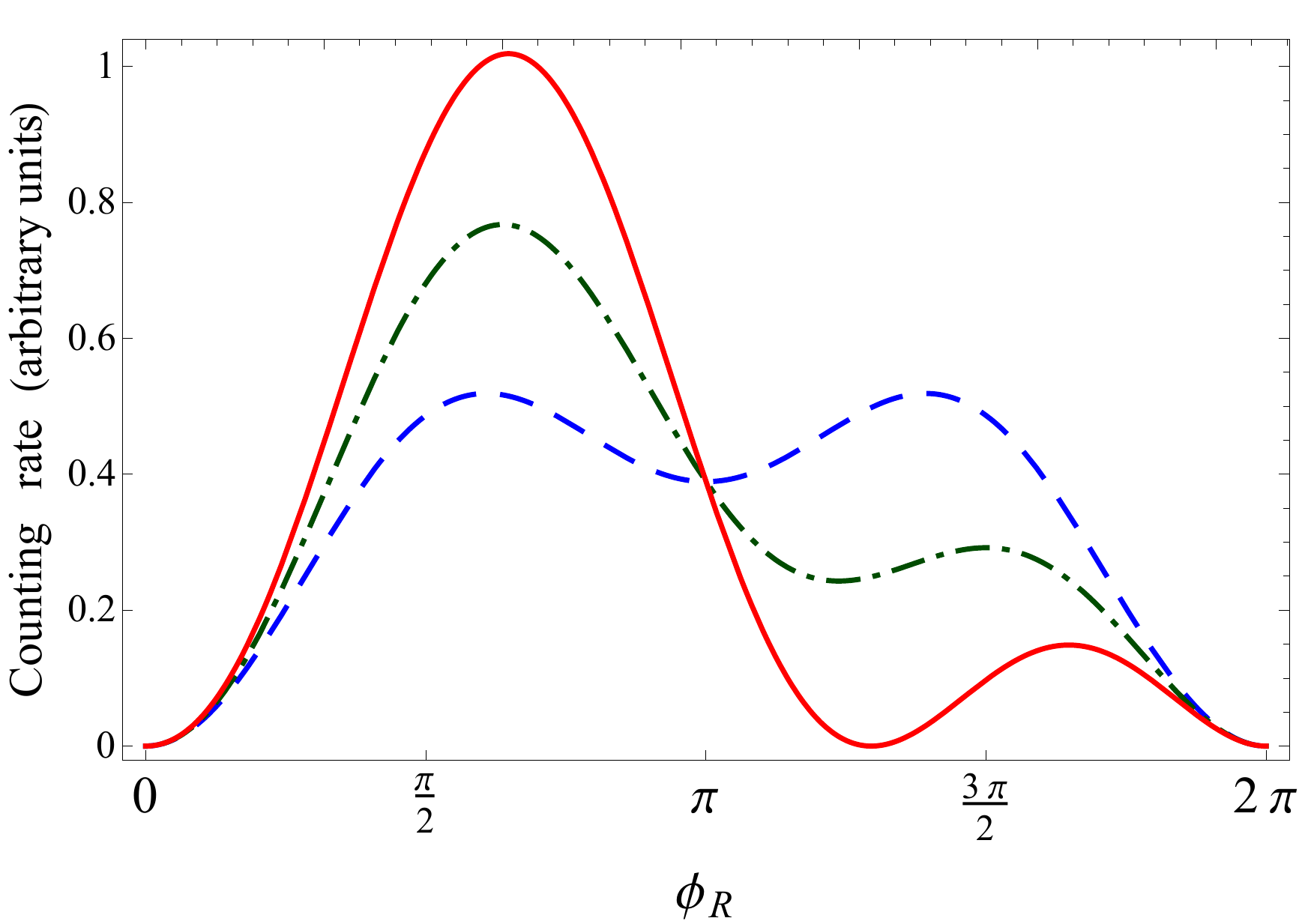}
\caption{(color
  online). Counting rate corresponding to the simultaneous detection in the angular momentum state $| \psi^{\rm det} \rangle$. Blue curve (dashed line), green curve (dashed-dot line) and the red curve (full line) shows the signal obtained for an initially incoherent ($\lambda=0$), partially coherent  ($\lambda=1/2$) and fully coherent ($\lambda=1$) density matrix respectively. The left Zeeman phase shift has been fixed to $\phi_L=\pi/2$. }
\label{fig2}
\end{figure}

Fig.~\ref{fig2} shows the profile of the detection signal as a function of the phase $\phi_R$ imprinted by the right phase object onto the atomic waves, for different values of the coherence parameter $\lambda$ in the initial density matrix~(\ref{eq:initial density matrix}), and for a fixed value $\phi_L= \pi/2$ of the atomic phase provided by the left phase object. One notes that for a value of $\phi_R=0,\pi$, the detection signal becomes insensitive to the initial coherence parameter $\lambda$. This fact, obvious from Eq.(\ref{eq:detection signal}), can be interpreted as follows. In either case, one can show that the evolved quantum states $\hat{U} | \psi_0 \rangle $ and $\hat{U} | \psi_1 \rangle $ are orthogonal, which means that an one could with a suitable measurement, know with certainty the original angular momentum state of the molecule (either $| \psi_0 \rangle$ or $| \psi_1 \rangle$). In this specific case, the apparatus would destroy an eventual initial spin-coherence in the density matrix. For $\phi_R=0$ or $\phi_R=\pi$, the atomic waves carrying an angular momentum of $| m_x = \pm 1 \rangle$ suffer an identical phase shift in the right phase object. This shows the key role played by the SGIs in the system, which by mixing the angular momentum states, permits the observation of quantum interferences. When the phases $\phi_{L,R}$ are in the vicinity of $\phi_{L,R} \simeq \pm \pi/2 (2\pi)$, the overlap between the quantum states  $ \hat{U} | \psi_0 \rangle$ and $ \hat{U} | \psi_1 \rangle $ is maximum.

This justifies our choice $\phi_L=\pi/2$ for the phase imprinted by the left magnetic field gradient, which maximizes the desired interference pattern. For this value, the detection profiles are indeed qualitatively very different  for a pure initial statistical mixture [$\lambda=0$] and for a pure initial quantum state [$\lambda=1$], and is thus well suited to investigate the initial spin coherence. Precisely, for an initially incoherent density matrix, one obtains a superposition of two independent peaks of same height centered around $\phi_R \simeq \pi /2 , 3 \pi /2$. In contrast, for a fully coherent state, one observes a strong enhancement of the first peak  ($ \phi_R \simeq \pi/2$), whereas the second peak is strongly reduced and  displaced to the right ($ \phi_R > 3 \pi/2$). This can be understood in terms of constructive interferences on the first peak combined with destructive interferences on the second peak.
 For a partially coherent initial density matrix ($\lambda=1/2$), one observes a partial suppression of the second peak. Indeed, by estimating the asymmetry between the two peaks around the detection angle $\phi_R=\pi$, one can differentiate between an incoherent and coherent initial density matrix. 

Finally, in order to verify the robustness of our result we have introduced uncertainty in the phases $\phi_L$ and $\phi_R$, once they depend on all critical physical quantities for the accomplishment of the experiment. We let the phases fluctuate up to 15\% and even with this degree of uncertainty one  can still distinguish the interference pattern arising from an initially completely  incoherent density matrix (statistical mixture) from the one resulting from a pure state. These are associated to the extreme cases corresponding respectively to a completely spin-incoherent and completely spin-coherent molecular dissociation. In fact, considering the behavior of the peaks one will even be able to note an intermediate coherence as well.   
 
 To conclude, we have shown and analysed an atom interferometer configuration able to address the spin coherence of a molecule dissociation process, starting from a quantum state of null total angular momentum and with fast and metastable atoms. We have shown that a configuration based on two symmetrically disposed magnetic field gradients, combined with a set of polarizers and analysers, yields an asymmetric atomic interference pattern for a spin-coherent dissociation. The role played by the SGIs is to appropriately shuffle the angular momentum states of the two identical particles propagating in the system, enabling quantum interferences from initially orthogonal quantum states. This experimental setup enables one to clearly distinguish a spin-coherent from a spin-incoherent fragmentation, and may allow to estimate the density matrix purity by measuring the asymmetry of the detection signal when the magnetic field of one of the phase objects is varied. This setup could be used to assess the coherence of various molecular dissociation processes as well as their relevance for twin-atom interferometry experiments.\\
 
\noindent Emails: (a)crenato@if.ufrj.br,\\  (b)jalbert@if.ufrj.br,  (c)impens@if.ufrj.br


\begin{thebibliography}{99}

\bibitem{Einstein35} EINSTEIN A., PODOLSKY B., and ROSEN N., \textit{Phys. Rev.}, \textbf{47} (1935) 777.

\bibitem{Schrodinger35} SCHR\"ODINGER E., \textit{Mathematical Proceedings of the Cambridge Philosophical Society}, \textbf{31} (1935)  555. 

\bibitem{Bohr58} BOHR N., ``Discussion with Einstein on epistemological problems in atomic physics", {\it Atom Physics and Human Knowledge}, (John Wiley \& Sons, 1958).

\bibitem{Bohm89} BOHM D., \textit{Quantum theory}, (Eglewood Cliffs, NJ: Prentice-Hall, 1951); re-edited in 1989 as {\it Quantum Theory}, (Dover Publications, Inc, New York, 1989). 

\bibitem{Clauser69} CLAUSER J. F., HORNE M. A., SHIMONY A., and HOLT R. A., \textit{Phys. Rev. Lett.}, {\bf 23} (1969) 880.

\bibitem{Bell} BELL J. S., Physics, {\bf 1} (1964) 195.

\bibitem{Aspect} ASPECT A., GRANGIER P., and ROGER G., \textit{Phys. Rev. Lett.}, \textbf{49} (1982) 91; ASPECT A., DALIBARD J., and ROGER G., {\it ibid}, \textbf{49} (1982) 1804.

\bibitem{Schwinger88} ENGLERT B.-G. , SCHWINGER J., and SCULLY M. O., \textit{Found. Phys.}, {\bf 18} (1988) 1045;  SCHWINGER J., SCULLY M. O., and ENGLERT B.-G. , \textit{Z. Phys. D}, {\bf 10} (1988) 135; SCULLY  M. O.,  ENGLERT B.-G. , and SCHWINGER J., \textit{Phys. Rev. A}, {\bf 40} (1989) 1775.

\bibitem{Caldeira06} DE OLIVEIRA T. R., and CALDEIRA  A. O., \textit{Phys. Rev. A}, \textbf{73} (2006) 042502.

\bibitem{Fry00} FRY E. S. and WALTHER T. , \textit{Adv. Atom. Mol. Opt. Phys.}, {\bf 42} (2000) 1.

\bibitem{CroninRMP09} CRONIN  A. D.,  SCHMIEDMAYER J., and PRITCHARD D. E., \textit{Rev. Mod. Phys.}, \textbf{81} (2009) 1051.

\bibitem{Deng99}  DENG L., HAGLEY E. W., WEN J., TRIPPENBACH M., BAND Y., JULIENNE P. S., SIMSARIAN J. E., HELMERSON K., ROLSTON S. L., and PHILLIPS W. D., \textit{Nature}, \textbf{398} (1999) 218.

\bibitem{WestbrookNature07}  SCHELLEKENS M., HOPPELER R., PERRIN A., GOMES J. V., BOIRON D., ASPECT A., and WESTBROOK C. I., \textit{Nature}, \textbf{445} (2007) 402.

\bibitem{Hagley97} HAGLEY E., MA\^ITRE X., NOGUES G., WUNDERLICH C., BRUNE M., RAIMOND J. M., AND HAROCHE S., \textit{Phys. Rev. Lett.}, \textbf{79} (1997) 1.

\bibitem{Tanabe09} TANABE T., ODAGIRI T., NAKANO M., SUZUKI I. H., AND KOUCHI N., \textit{Phys. Rev. Lett.}, \textbf{103} (2009) 173002.

\bibitem{Bucker11} B\"UCKER R., GROND J., MANZ S., BERRADA T., BETZ T., KOLLER C., HOHENESTER U., SCHUMM T., PERRIN A. AND SCHMIEDMAYER J., \textit{Nature Phys.} \textbf{7} (2011) 608.


\bibitem{Zeilinger}  KWIAT P. G.,  MATTLE K., WEINFURTER H., ZEILINGER A., SERGIENKO A. V. , and SHIH Y., \textit{Phys. Rev. Lett.}, {\bf 75} (1995) 4337.

\bibitem{Robert-Baudon} ROBERT J., MINIATURA CH., GORCEIX O., LE BOITEUX S., LORENT V., REINHARDT J. AND BAUDON J., \textit{J. Phys. II France}, \textbf{2} (1992) 601; MATHEVET R., BRODSKY K., BAUDON J., BROURI R., BOUSTIMI M., VIARIS DE LESEGNO B., AND ROBERT J. , \textit{Phys. Rev. A}, \textbf{58} (1998) 4039; VIARIS DE LESEGNO B., KARAM J.C., BOUSTIMI M., PERALES F., MAINOS C., REINHARDT J., BAUDON J., BOCVARSKI V., GRANCHAROVA D., PEREIRA DOS SANTOS F., DURT T., HABERLAND H., AND ROBERT J. \textit{Eur. Phys. J D}, \textbf{23} (2003) 25.

\bibitem{SGA07}  PERALES F., ROBERT J., BAUDON J. , and M. DUCLOY, \textit{Eur. Phys. Lett.}, \textbf{78} (2007) 60003.

\bibitem{Medina11} 
MEDINA ALINE, RAHMAT G., DE CARVALHO C. R., JALBERT GINETTE, ZAPPA F., NASCIMENTO R. F., CIREASA R., VANHAECKE N., SCHNEIDER IOAN F., DE CASTRO FARIA N. V. and ROBERT J., \textit{J. Phys. B: At. Mol. Opt. Phys.}, \textbf{44} (2011) 215203.

\bibitem{Medina12} 
MEDINA ALINE, RAHMAT G., JALBERT GINETTE, CIREASA R., ZAPPA F.,  DE CARVALHO C. R., DE CASTRO FARIA N. V., and ROBERT J., \textit{Eur. Phys. J. D}, \textbf{66} (2012) 134.

\bibitem{Robert13} 
ROBERT J., ZAPPA F., DE CARVALHO C. R., JALBERT GINETTE, NASCIMENTO R. F., TRIMECHE  A., DULIEU O., MEDINA ALINE, CARVALHO CARLA, AND DE CASTRO FARIA N. V., \textit{Phys. Rev. Lett.}, {\bf 111} (2013) 183203.

\bibitem{Lamb} LAMB W. E. JR. and RETHERFORD R. C., Phys. Rev. 79 (1950) 549 (see sections 16 and 17).

\bibitem{Robert89} ROBERT J., MINIATURA CH., PERALES F., VASSILEV G., BOCVARSKI V., REINHARDT J., BAUDON J. AND LORENT V,~\textit{Euro Phys. Lett.}, {\bf 9} (1989) 651.
 
 
 \bibitem{tttscheme} BORD\'{E} C. J., \textit{Gen. Rel. Grav.}, \textbf{36} (2004) 475; ANTOINE C., \textit{App. Phys B}, \textbf{84} (2006) 585.


\bibitem{ABCD}  BORD\'{E} C. J., \textit{Metrologia}, \textbf{39} (2002) 435; IMPENS F. and BORD\'{E} C. J., \textit{Phys. Rev. A}, \textbf{79} (2009) 043613.

\bibitem{Varshalovich}  VARSHALOVICH D. A. and MOSKALEV A. N., \textit{Quantum Theory of Angular Momentum}, (World Scientific Publishers, Singapore, 1988).

\bibitem{Siebbeles91}  SIEBBELES L. D. A., SCHINS J. M., VAN DER ZANDE W. J., AND J. A. BESWICH J. A., \textit{Chem. Phys. Lett.}, \textbf{187} (1991) 633.

\bibitem{Kwiat04} PETERS N. A., WEI T.-C. , AND KWIAT P. G., \textit{Phys. Rev.  A}, \textbf{70} (2004) 052309.

\end{thebibliography}
\end{document}